\def\year{2022}\relax
\documentclass[letterpaper]{article} 
\pdfoutput=1
\usepackage{aaai22}  
\usepackage{times}  
\usepackage{helvet}  
\usepackage{courier}  
\usepackage[hyphens]{url}  
\usepackage{graphicx} 
\usepackage{natbib}  
\usepackage{caption} 
\DeclareCaptionStyle{ruled}{labelfont=normalfont,labelsep=colon,strut=off} 
\usepackage{algorithm}
\usepackage{algorithmic}
\usepackage{newfloat}
\usepackage{listings}
\floatstyle{ruled}
\newfloat{listing}{tb}{lst}{}
\floatname{listing}{Listing}
\title{Evaluating TCFD Reporting: A New Application of Zero-Shot Analysis to Climate-Related Financial Disclosures}
\author{
    Alix Auzepy, \textsuperscript{\rm 1}
    David Lenz, \textsuperscript{\rm 1}
    Elena Tönjes, \textsuperscript{\rm 1}
    Christoph Funk \textsuperscript{\rm 2}
}
\affiliations{
    \textsuperscript{\rm 1}Justus Liebig University Giessen\\
    Faculty of Economics and Business Studies\\
    Licher Strasse 64\\
    35394 Giessen, Germany\\
    
    \textsuperscript{\rm 2}Justus Liebig University Giessen\\
    Centre for International Development and Environmental Research (ZEU)\\
    Senckenbergstrasse 3\\
    35390 Giessen, Germany\\
    
    Alix.Auzepy@wi.jlug.de, 
    David.Lenz@wi.jlug.de,
    Elena.Tönjes@wi.jlug.de,
    Christoph.Funks@wi.jlug.de
    
}

\begin{document}

\maketitle

\begin{abstract}
We examine climate-related disclosures in a large sample of reports published by banks that officially endorsed the recommendations of the Task Force for Climate-related Financial Disclosures (TCFD). In doing so, we introduce a new application of the zero-shot text classification. By developing a set of fine-grained TCFD labels, we show that zero-shot analysis is a useful tool for classifying climate-related disclosures without further model training. Overall, our findings indicate that corporate climate-related disclosures grew dynamically after the launch of the TCFD recommendations. However, there are marked differences in the extent of reporting by recommended disclosure topic, suggesting that some recommendations have not yet been fully met. Our findings yield important conclusions for the design of climate-related disclosure frameworks.\\
\end{abstract}

\section{Introduction}

Published in 2017, the recommendations of the Financial Stability Board's Task Force on Climate-related Financial Disclosures (TCFD) have been described by the Government of the United Kingdom as “one of the most effective frameworks for companies to analyse, understand and ultimately disclose climate-related financial information”\cite{UK2021}. 

The TCFD recommendations, which have been formally endorsed by more than 4,000 companies worldwide to date, are a set of voluntary disclosure guidelines aimed at providing consistent climate-related information to investors and other key company stakeholders \cite{Friederich2021}. Compared to other reporting frameworks (e.g. Carbon Disclosure Project, Global Reporting Initiative), a particular focus of the TCFD recommendations is on disclosing information on the integration of climate-related risks into risk management processes, control structures and key aspects of business operations \cite{Beyene2022b}. Overall, the recommendations are structured around four broad disclosure categories (Governance, Strategy, Risk Management, Metrics and Targets) and 11 underlying recommended disclosures. For an overview of the TCFD recommendations, see Figure \ref{figure:TCFDrecos} in the Appendix. 

Since reliable information on climate-related risk exposures is critical for making informed investment decisions and appropriately pricing risks, an increasing number of investors have been exerting pressure on companies to issue reports that include comprehensive climate-related financial disclosures \cite{Krueger2020}. In addition, several countries, including the UK and Switzerland, have taken steps to make TCFD reporting mandatory for large companies in their jurisdictions. From a company perspective, the disclosure of climate-related information often signals awareness and preparedness for climate-related issues, while the absence of disclosure may, on the contrary, indicate that such issues are not being addressed by the company \cite{Sullivan2012, Bingler2022}. Illustrating these arguments, \citet{Jung2018} and \citet{Subramaniam2015} show that firms are more likely to integrate risks associated with climate change into their overall risk management when they also disclose information about such risks.

Against this background, it is surprising to find that research on climate-related disclosures is very sparse. What is more, prior studies have shown rather contrasted results with regard to the ability of the TCFD guidelines in delivering high-quality and material information due to several reasons, including greenwashing as well as a lack of standardization, quantitative data and transparency. For example, \citet{Bingler2022} investigated climate-related disclosures related to the four core disclosure pillars based on a sample of 818 TCFD-supporting firms from 2015 until 2020. The authors came to the conclusion that climate-related reporting is associated with selective disclosure, suggesting that firms disclose climate-related information primarily on the least material TCFD disclosure categories, i.e. Governance and Risk Management. \citet{Ding2022} analyzed how carbon emissions affect voluntary climate-related disclosures based on TCFD. Their results show that firms with higher levels of carbon emissions disclose more climate-related information. In particular, they find that carbon emissions drive disclosure at the category level for the Strategy, Risk Management and Metrics and Targets pillars, but not for the Governance pillar. 

So far, these studies have left a detailed analysis of the recommended disclosures largely untouched. In particular, the existing literature focuses on the quantity of information at the broad TCFD category level, rather than on the content \cite{Mehra2022}. A deeper analysis is crucial, however, as the content and materiality of the disclosures vary according to the sector of the company concerned. In particular, the TCFD recommendations are primarily aimed at financial institutions. As climate change affects the credit risk of different types of assets and poses the risk of stranded assets, high exposure to climate change by financial institutions could also increase the risk of financial instability \cite{Beyene2021}. Thus, the TCFD recommendations place great emphasis on the disclosure of concentrations of carbon-related assets in the financial sector \cite{FSB2015}. Furthermore, the TCFD published supplemental guidance for the financial sector and encourages banks to provide the metrics used to assess the impact of climate-related risks on their lending and other financial activities.

In this paper, we attempt to address this gap in two important ways. First, we provide new insights into the state of disclosure regarding the underlying TCFD recommendations by analyzing a sample of 3.335 reports published by TCFD-supporting banks between 2010 and 2021. We identify TCFD supporters by retrieving all banks that have publicly declared their support to TCFD and are listed as official supporters on TCFD's website. We focus explicitly on banks for the reasons mentioned before. Overall, we contribute to the literature on the use of natural language processing (NLP) in the context of climate-related financial disclosures by introducing the zero-shot text classification as a new method for systematic and automated extraction of textual information from large amounts of reports. In addition, we develop a set of fine-grained labels that allow us to leverage information on the TCFD recommendations beyond the four core pillars. 

The zero-shot classification returns label probabilities for each extracted text sequence. This can therefore be interpreted as the probability that the corresponding sequence matches the label, or in other words, the probability that the text sequence deals with the topic addressed by the label. We assume that higher label probability is a proxy for disclosure quality because higher label probability indicates that the semantics of the text sequences are likely to match the semantics of the labels at hand. Thus, when text sequences precisely and explicitly address a topic expressed in a label, they are associated with a high label probability in the context of zero-shot text classification. To some extent, a higher label probability is also an indicator of how much is disclosed about a particular topic, since labeling is associated with a higher probability if a text sequence discusses the topic in question in detail. 

Compared to other language models, the zero-shot approach has the crucial advantage of being able to classify sentences using labels for which it has received no prior training. A weakness of algorithms trained to automatically identify and classify climate-related content is that such models require an extensive training set of human-labeled sentences. In particular, manual labeling of sentences is not only time-consuming, but can also be error-prone. Therefore, for quality and consistency reasons, highly-trained and specialized “labelers” are required, which can also make the labeling process costly. Furthermore, the more classes (or categories of labels) to be included into the classification scheme of the model, the more labeled data is needed to ensure that each class comes with a reasonable amount of examples attached to it, which can be a limiting factor in some scenarios. As our method does not require any labeled training data, it also does not impose any restrictions on the number of classes, which allows us to perform a more detailed analysis of the underlying TCFD recommended disclosures. Additionally, the TCFD recommendations are well-suited for the zero-shot analysis, as they provide us with an already-existing framework and semantics \cite{Ding2022}. In total, we develop 14 fine-grained labels designed to capture the most central aspects of the TCFD recommendations for banks using similar semantics. 

Our paper yields the following sets of findings. First, we investigate the level of disclosure at category level. Specifically, we examine the mean probabilities that the text sequences in our sample relate to one of the four core TCFD categories. We find that the mean probabilities relating to the general labels (i.e., Governance, Strategy, Risk Management, and Metrics and Targets without explicit mention of climate) remain stable over the sample period from 2010 to 2021, while we observe an increase in all of the probabilities for the climate-related labels at category level (i.e., climate-related Governance, climate-related Strategy, climate-related Risk Management and climate-related Metrics and Targets) over the same time period. In particular, we report that the disclosures pertaining to the labels “climate-related Strategy” and “climate-related Metrics and Targets” grew particularly dynamically, reaching mean probabilities of 22\% and 20\% respectively, in 2021. At first sight, this might suggest that TCFD-supporting banks are more likely to focus on the strategic relevance of climate-related risks to their business or the development of climate-related business strategies with corresponding targets.

However, one way to more accurately assess the quality of TCFD recommendation disclosures and their corresponding implementation by companies might be to look not only at the broad TCFD categories, but rather to examine corporate reporting on recommended disclosures within each broad category. To the best of our knowledge, the existing literature on TCFD has so far only investigated reporting at the category level. Since text sequences in corporate reports oftentimes correspond to several categories at once, the results from such analyses might lead to diverging conclusions. In a second step, we therefore analyze the mean probabilities associated with our fine-grained labels, which cover the underlying recommended disclosures. Our results indicate that there is considerable variation in disclosures, including within each TCFD category. In the Strategy area, which is the most comprehensive one and contains several specific recommended disclosures for banks, we find that label probabilities are lower for disclosures related to financing and investment activities for carbon-intensive industries such as the fossil fuel industry. Similarly, TCFD-supporting bank appear less likely to explicitly address the use of climate-related scenario models in their disclosures. 

Under Metrics and Targets, we find that the incorporation of climate-related performance metrics into remuneration policies is associated with a lower label probability compared to labels related to carbon footprints and emissions reduction targets. In the Governance area, the TCFD-supporting banks appear to report to similar levels on the board's responsibility for overseeing climate-related issues and the management's role in assessing and managing climate-related issues. However, maximum values for board oversight are higher, indicating a higher quality of reporting in some reports on the role of the board in overseeing climate-related issues. 

The remainder of this paper is organized as follows. First, we present our data, followed by our methods and model performance evaluation. Our results section is twofold. In the first part, we present the results of the zero-shot classification at the category level. In the second part, we analyze the results for the fine-grained labels covering the TCFD recommended disclosures. The results are summarized and discussed in the last section. 

\section{Data}\label{sec:data}

We apply the zero-shot classification to a sample of 3,335 hand-collected reports between 2010 and 2021. Due to the large differences between the homepages of the TCFD-supporting banks in our sample, a fully automated scraping approach is not possible. As a first step, we retrieve all the names of TCFD-supporting banks from the TCFD website by filtering them according to the industry categories available on the website: “banks”, “central banks” and “capital markets”. After removing the banks that could not be identified or for which reports were not available online, we are left with 188 TCFD-supporting banks. Table \ref{DescriptiveI} presents the bank level data.

As a next step, we categorize the banks in our sample according to the region of their headquarters and their total asset size.   In our analysis, banks with total assets greater than USD 500 billion are considered “large”, whereas banks with total assets between USD 50 billion and USD 500 billion are “medium”, and banks with less than USD 50 billion are “small”. Interestingly, almost half of our sample consists of banks from the Asia-Pacific region. European banks account for one third and North American banks for about 10\% of our sample. The majority of banks in our sample are mid-sized banks with total assets between USD 50 billion and USD 500 billion. 

\begin{table}[!ht]
\begin{center}
\setlength{\tabcolsep}{4pt}
\begin{scriptsize}
\caption{Size and region of TCFD-supporting banks}
\label{DescriptiveI}
\begin{tabular}{lrrrr} 
\hline
Region                & Large                & Medium  & Small                & $\sum$                \\
\hline
Asia Pacific          & 15                   & 51                   & 24                   & 90                  \\
Europe                & 23                   & 26                   & 17                   & 66                   \\
Latin America         & 0                    & 4                    & 3                    & 7                    \\
Middle East \& Africa & 0                    & 3                    & 2                    & 5                    \\
North America         & 9                    & 8                    & 3                    & 20                   \\ \hline
$\sum$        & 47                    & 92                    & 49                   & 188                 \\
\hline
\end{tabular}
\end{scriptsize}
\end{center}
\end{table}

We follow the approach in \citet{Bingler2022} and collect available bank reports for the period 2010 to 2021 to capture textual data both before and after the publication of the TCFD recommendations in June 2017. The reports are classified according to the following categories: annual reports, CDP reports, corporate governance reports, integrated reports, remuneration reports, sustainability reports, and TCFD reports. We do not rely solely on TCFD reports, as most TCFD supporters do not publish stand-alone TCFD reports for their climate-related disclosures, but rather tend to integrate key information into their annual and sustainability reports. This is consistent with the TCFD recommendations, which stress that climate-related disclosures should be included in “mainstream (i.e., public) annual financial filings” \cite{TCFD2017a}. The majority of reports in our sample consists of annual and sustainability reports.

After parsing the reports to ensure they are in a suitable raw text format for the zero-shot classification, we are left with a total sample of 3,335 bank reports, as illustrated in Table \ref{DescriptiveII}. Since extracting information on climate disclosures at a large scale is associated with immense manual work, the zero-shot allows us to examine a comparatively large sample of reports compared to previous TCFD-related studies. For example, \citet{Ding2022} apply computerized textual analysis to a sample of 140 reports from TCFD signatories, \citet{Demaria2020} analyze a sample of French CAC 40 firms from 2015 until 2018. 

\small
\begin{table}[!ht]
\begin{scriptsize}
\begin{center}
\setlength{\tabcolsep}{3pt}
\caption{Sample composition}
\label{DescriptiveII}
\begin{tabular}{lrrr}  
    \hline
Report Category             & \# of reports& Average pages & Average \# of sentences \\
    \hline
Annual Report               & 1869                                  & 207.98                            & 2711.21                                 \\
CDP Report                  & 75                                    & 63.43                             & 699.79                                  \\
Corporate Governance Report & 148                                   & 69.44                             & 1014.25                                 \\
Integrated Report           & 183                                   & 163.98                            & 2354.95                                 \\
Remuneration Report         & 83                                    & 36.88                             & 494.42                                  \\
Sustainability Report       & 896                                   & 81.01                             & 1158.54                                 \\
TCFD Report                 & 81                                   & 36.68                             & 544.37                                  \\\hline
             &                $\sum = 3335$                       & $\bar{x}$ = 94.20                             & $\bar{x}$= 1282.50     \\ 
    \hline
\end{tabular}
\end{center}
\end{scriptsize}
\end{table}

\normalsize
\subsection{Parsing PDFs} 

All of the reports used in our analysis are in PDF format. Converting textual information contained in PDF documents into a suitable format for further NLP analysis is not as trivial as analyzing textual information stored in CSV or TXT files. In this paper, we apply a layout-parsing model, which is able to detect and extract actual text from PDF documents. In the context of our analysis, we include the actual text from the reports and deliberately omit the text from tables and graphs, which not only increases the quality of our data, but also saves computation time. 

Our parsing model is based on Visual-Layout (VILA) groups introduced by \citet{Shen.01.06.2021}. VILA is able to convert textual data into groups of tokens (text lines or blocks) and to assign a layout tag to these tokens. There are several variants of VILA, called H-VILA (Visual Layout-guided Hierarchical Model) and I-VILA (Injecting Visual Layout Indicators). After several trials, we choose the H-VILA block variant trained on grotoap2 using the layoutLM model \cite{Xu.2020b} as a base model since it delivered the best extraction and tokenization results. The output consists of the extracted text as groups of tokens together with the corresponding layout tags. Depending on the training set, the layout tags can be figures, body content, abstract and title. For our analysis, we keep the parts tagged as body content and abstract. 

\subsection{Zero-Shot Classification}
\label{zeroshot}

A widely used and important NLP task is text classification \cite{belinkov2019analysis}. Text classification is used to organize and analyze very large amounts of textual data by assigning classes, or so-called “labels”, based on the topic of individual text sequences, which may consist of sentences, paragraphs, or entire pages. Typically, simple neural network models, but also more complex language models with classification end-layers are used to perform this text classification. To this end, base models (e.g., BERT or BART) are pre-trained on an extensive amount of textual data in order to learn semantics \cite{Mehra2022}. They can afterwards be used for several NLP tasks and thus be trained on specific tasks in the fine-tuning stage. To this end, different combinations of several pre-trained language models (the base model) and task-specific end-layers can be used. 

One way to fine-tune a base model is to train it with labeled training data. Therefore, the accuracy of the models often also depends on the size and quality of the training data. Creating a training set for sentence classification has several drawbacks: First, manually labeling large amounts of text is very time consuming and requires a lot of manpower. Second, the labeling process must be repeated when classes need to be changed or new labels need to be added. Third, it can be difficult even for humans to assign the correct label to a sentence, since certain sentences may interpreted in different ways. Another drawback is that the humans labeling the data set are often biased, which makes it difficult to get a representative training set \cite{Beltagy.}. Finally, since the training data cannot be the same as the data used in the actual analysis, finding enough suitable training data can be a challenge. For example, in the case of an analysis of TCFD reports, the labeled reports that were used to train the model cannot be used afterwards. Additionally, since the number of TCFD reports published by banks is limited and the TCFD reporting by companies is also imperfect, this automatically leads to a limitation in finding high-quality training data. 

Due to the drawbacks mentioned above, we employ in this paper a zero-shot text classification model introduced by \citet{Davison.2020}. The model is able to classify text sequences based on the semantics of the input sequences and the labels without further training. A simplified structure of our model architecture is shown in Figure \ref{fig:BART} in the Appendix. As a base model for the zero-shot classification, we use BART, which is pre-trained on roughly 160GB of text from the English Wikipedia and BookCorpus dataset in order to “understand" the semantics of texts \cite{Mehra2022}. Since the TCFD recommendations do not include specific financial language, but rather general semantics, we consider such base model to be well-suited for our analysis. Furthermore, since the reports in our sample are also in the English language, we benefit from the fact that the model was pre-trained on a very large English language corpora. In contrast to other models such as BERT, the BART model \cite{Lewis2019} not only makes use of the sequence-to-sequence translation architecture with bidirectional encoders (BERT), but also uses a left-to-right autoregressive decoder (GPT model) and is therefore a mixture of both. In combination with the zero-shot text classification, BART as a base model demonstrates high performance results \cite{Davison.2020}.

The specific NLP task used for our zero-shot classification consists of a multi-natural language inference (MNLI) \cite{Davison.2020}. The zero-shot text classification embeds the sentences of a text (a sequence of words) and the labels themselves into the same latent space. In such a latent space, the distance between the sentence and the label can be computed. The closer the label is to the sentence, the higher the probability that the label matches the sentence. In the context of the zero-shot text classification, labels are therefore used to measure the probability that a text sequence addresses one (or more) labeled topics.

While the zero-shot text classification yields probabilities from 0 to 100\% for each label, the ClimateBERT model used in \citet{Bingler2022} does not return probabilities as output, but a binary output where one label is considered as true and all others as false. For example, the classification output of one governance-related paragraph could be 1 for the label Governance, and 0 for all other labels, i.e. 0 for Strategy, 0 for Risk Management etc. In contrast to our approach, the authors measure the proportion of TCFD-related content in corporate reports by summing up the results for all labels and putting them in relation to the number of paragraphs. 

In contrast, our text classification model treats text sequences as premises and labels as hypotheses. It tokenizes them and uses the underlying language to embed both the sentence and the label. The model runs both through the pre-trained MNLI layer. The MNLI end-layer is a simple fully-connected neural network where the output is a vector of logit scores for three outcomes: “neutral”, “contradiction” and “entailment” \cite{Lewis2019, Davison.2020}. Hence, the hypothesis is tested against the premise and the result can be a classification as entailment, a contradiction, or neutral. The score for “neutral” is discarded and a softmax function is applied to the “contradiction” and “entailment” scores in order to be able to interpret them as a probability on a scale from 0 to 100\%. In our analysis, the scores shown are for the “entailment” only. They can therefore be interpreted as the probability that the corresponding sequence matches the label, or in other words, the probability that the label is true. In a next step, the fine-tuned end layer can be used for zero-shot classification without any further training. 

Finally, due to the fact that the TCFD recommendations are interconnected and that a sentence can fit into several recommended disclosures at once, we employ a multi-label approach. Thus, we do not force the zero-shot text classification to return probabilities that add up to one as in the single label approach. Instead, we choose an approach where the model is able to assign probabilities from 0 to 1 for each label (multi-label approach) to account for the fact that a sentence could fit into more than one label. Consequently, the results for all labels for each sentence do not add to one.

\section{Label Evaluation}\label{sec:robustness}

The zero-shot text classification does not require specific training with pre-labeled data. Hence, we cannot validate the model in the usual way where we would split a data set into a training set and a test set and validate our model using the test set. Nevertheless, we can still perform a battery of evaluation tests to assess the text sequence recognition and classification performance of our zero-shot text classification model.

We create a new test data set by manually collecting sentences that are consistent with the TCFD recommendations. The data set consists of sentences from the TCFD good practice handbook \cite{CDSBa, CDSBb}, which contains examples of best practice disclosures selected by the TCFD. Furthermore, we extract sentences from TCFD reports of companies that are not in the banking sector. Finally, we also use sentences from the training repository made available by \citet{ClimateBERT}. We choose a multi-label approach in applying the zero-shot classification to account for the fact that a text sentence may correspond to several labels. Consequently, the results for all labels for each sentence do not add to one.

We employ the fine-grained labels provided in Table \ref{table2}. Furthermore, we include the label “none” to the classification task. The purpose of this label is to capture the non-climate-related text, i.e. the text sequences that do not fit any of our fine grained labels. It also ensures that the labels are not randomly assigned by the zero-shot, but that the assignment is really based on the semantics of the text sequences. In order to test the performance of our model with regards to the “none” label, we use sentences labeled as “none” in the training repository mentioned above, as well as additional sentences labeled by us as “none”.

The results of the zero-shot text classification using our fine grained labels are presented in Figure \ref{fig:our_labels} in the Appendix. The sentences manually annotated are represented on the x-axis, while the results from our zero-shot analysis based on climate-related labels are on the y-axis. The results represent the mean probability returned by the zero-shot model that the sentences in the column deal with the topic of the label in the row. The darker the entries, the higher the likelihood classified by the model. In the case of an optimal zero-shot text classification, the entries in the diagonal would be the darkest.

These results show that our model provides satisfactory results by assigning high probabilities to the relevant labels. At the same time, we also observe that labels ST.1.1 (Climate-related transition risks), ST.1.3 (Material financial impact of climate-related issues), ST.1.7 (Resilience of the bank's strategy), RM.1.1 (Processes to identify, assess and manage climate-related risks and integrate them into overall risk management), and RM 1.2 (Relationship between climate-related risks and financial risks) performed worse than other labels in terms of text sequence recognition. For example, for the labels RM.1.1 and RM1.2, not only those we labeled as risk management text sequences, but also several sentence groups were assigned these labels with a probability of 60\% or more.

There could be several reasons for these results. First, a potential reason could be that the labels are too broadly defined and that the topics addressed are too abstract for zero-shot text classification. For example, concepts such as “resilience” or “risk management processes” can be described in many different ways and could therefore be partially reflected in many text sequences. Second, the topics of the TCFD recommendations are often closely connected, so that text sequences can often fit several labels at once. For example, the only label which does not have the highest probability for its own sentences is label MT.1.1 (Carbon footprint, direct and indirect greenhouse gas emissions), where a higher probability was assigned by the zero-shot to MT.1.3 (emissions reduction and carbon neutrality targets). Since these topics are closely related, a high value for both sets of sentences is not surprising. However, we have made a conscious decision not to change the labels in order to be as close as possible to the semantics of the TCFD recommendations. 

The probabilities for the “none” label are very low across all sentences. This suggests that the model does not label correctly by chance, but actually recognizes and incorporates the semantics of the labels. Moreover, sentences tagged as “none” were classified with a relatively low probability by the zero-shot text classification for nearly all labels. The label “climate-related transition risks” (ST.1.1) has a slightly higher probability of 37\% for the “none” sentences. This may be linked to the fact that this label encompasses many different types of risks, such as political and legal risks, technological risks, market risks, and reputation risks, all of which belong to the “transition risks” category. For some sentences describing these risks, the zero-shot classification may not directly identify the link to climate change. Furthermore, we also notice that more abstract labels such as RM.1.1 and RM.1.2 have higher values for “none” sentences as well. 

In addition to our graphical analysis, we evaluate the model performance by examining the overall F1 scores and the respective F1 scores of our labels, as illustrated in Table \ref{tab:scores}. We evaluate the model based on test data used previously in Figure \ref{fig:our_labels} (see Appendix) focusing on our fine-grained labels. Overall, our model obtains a micro F1 score of 0.60 and a macro F1 score of 0.57, which is satisfactory considering that we have 14 classes. We also find a large variance in performance across classes, with material financial impact of financial issues (ST.1.3) being the hardest to identify (F1-score of 0.34) and the incorporation of climate-related performance metrics into remuneration policies (MT.1.2) the easiest (F1-score of 0.78). This can be explained by the fact that financial material impact is a broader concept, and the task of identifying corresponding sentences may also be difficult even for humans. We also observe that our governance labels exhibit a comparatively high precision. In contrast, the labels pertaining to transition risks (ST.1.1), the relationship between climate-related risks and financial risks (RM.1.2) and emissions-reduction targets (MT.1.3) have a comparatively lower precision, suggesting a higher number of false positives. Additionally, our labels exhibit relatively high recalls, with the exception of material financial impact of financial issues (ST.1.3.) and financing and investment for carbon-intensive industries (ST.1.5). 

Altogether, our zero-shot text classification does not appear to assign probabilities purely by chance. The TCFD recommendations appear to be intertwined, which is an argument for the multi-label approach we use. We also find that zero-shot classification yields better results when labels are based on well-delineated and precisely defined concepts.

\begin{table*}[t]
\begin{scriptsize}
\centering
\caption{\label{tab:cleaningtab} Comparison of performance based on F1 scores}
\scalebox{1.1}{
    \begin{tabular}{lcccccccccccccc}
  \hline
        Label & GO.1.1 & GO.1.2 & ST.1.1 & ST.1.2 & ST.1.3 & ST.1.4 & ST.1.5 & ST.1.6 & ST.1.7 & RM.1.1 & RM.1.2 & MT.1.1 & MT.1.2 & MT.1.3 \\
        \hline
        Recall & 0.53 & 0.74 & 0.66 & 0.61 & 0.28 & 0.52 & 0.31 & 0.57 & 0.51 & 0.73 & 0.59 & 0.58 & 0.68 & 0.93\\
        Precision & 0.97 & 0.72 & 0.29 & 0.79 & 0.46 & 0.54 & 1.00 & 0.77 & 0.50 & 0.61 & 0.29 & 0.90 & 0.91 & 0.24\\
        F1-Score & 0.69 & 0.73 & 0.40 & 0.69 & 0.34 & 0.53 & 0.48 & 0.65 & 0.51 & 0.66 & 0.39 & 0.70 & 0.78 & 0.39 \\
        \hline
        \end{tabular}}
\centering
    \begin{raggedleft}The overall performance scores are: Micro F1-score: 0.6029, Macro F1-score: 0.5668, Weighted F1-score: 0.6281.
    \end{raggedleft}
    \label{tab:scores}
    \end{scriptsize}
\end{table*}

\section{Results}\label{sec:results}

\subsection{Climate-related disclosures by broad TCFD categories}\label{sec:Results1}

The main objective of the TCFD recommendations is to provide firms with guidance in disclosing consistent and decision-useful information for key stakeholders \cite{Bingler2022}. As climate-related disclosures are expected to reduce information asymmetries between reporting firms and stakeholders and to demonstrate a certain awareness from companies for climate-related topics \cite{Krueger2020, Jung2018, Ding2022}, we investigate the level of disclosure at category level. Specifically, we examine the probability that corporate reporting relates one of the four main TCFD pillars. 

Table \ref{tab:mean_per_year} presents the probabilities associated with the general labels “Governance”, “Strategy”, “Risk Management”, and “Metrics and Targets” as well as the probabilities for our labels “climate-related Governance” (GO.1), “climate-related Strategy” (ST.1), “climate-related Risk Management” (RM.1) and “climate-related Metrics and Targets” (MT.1), respectively. We intentionally include these two types of labels as a first step in comparing disclosures on general topics with disclosures on more specific climate-related topics.  

Several observations can be made based on these results: First, the mean label probabilities on general topics are higher than on specific climate-related topics. For example, the probability of reporting on “Governance” is higher, over the entire sample period, than the probability of reporting on “climate-related Governance” (GO.1). The same results hold for the other main categories. This finding seems plausible as we have extracted text sequences from different types of reports, including, for example, corporate governance reports and annual reports that discuss a variety of governance-related topics and do not focus exclusively on climate-related governance topics. Thus, the zero-shot text classification appears to make a good distinction between climate-related and non-climate-related textual data.

Second, we find that the mean probabilities relating to the general labels (without explicit mention of climate) remain stable over the sample period from 2010 to 2021, while we observe an increase in all of the probabilities for the climate-related labels after 2017. We find that the general label “Strategy” has the highest mean probability compared to the other labels. This label probability remains at 40\% (or 0.39\%) over time, while the label probability of "climate-related Strategy” grew particularly dynamically from 12\% in year 2010 to 22\% in year 2022. This indicates that while the probability that the text sequences in our sample dealt with “climate-related Strategy” was only 12\% for the reports pertaining to year 2010, it reached 22\% in year 2021. 

In addition, we find that the label “Metrics and Targets” has the second highest mean probability over the sample period compared to the other labels with a mean probability between 31\% and 34\%. Looking at the mean probability of “climate-related Metrics and Targets”, we find a comparatively more modest increase from 12\% for 2010 to 20\% for 2021, but still reaching a higher level than “climate-related Governance” and “climate-related Risk Management” in 2021. 

\begin{figure*}[t]
\centering
			\includegraphics[width=0.70\textwidth]{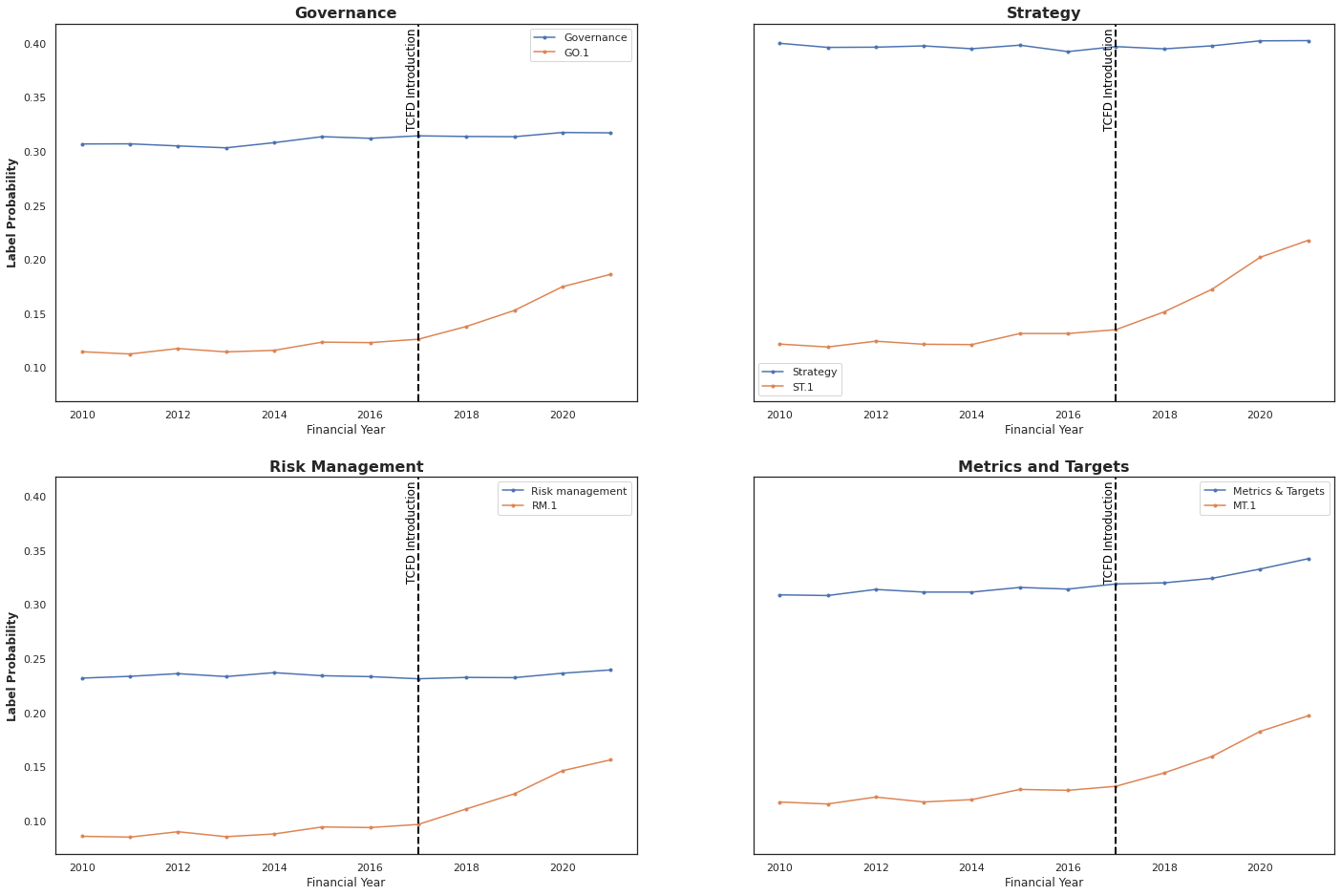}
			\caption{Climate-related disclosures by broad TCFD categories}\label{figure:onlycategorylabels}
\end{figure*}

To further examine climate-related disclosures at category level, we examine the trends in reporting before and after the publication of the TCFD recommendations in 2017. Figure \ref{figure:onlycategorylabels} illustrates these trends for all four TCFD categories over the sample period. Specifically, the blue lines represent the label probabilities of the general labels, while the orange lines illustrate the probabilities of the category labels with the explicit mention of “climate-related”, which are abbreviated as GO.1, ST.1, RM.1 and MT.1 in Table \ref{table2}. We find an increase in all climate-related label probabilities, which is particularly visible around 2017.

We find an increase in all climate-related label probabilities, which is particularly visible after 2017. For “climate-related Risk Management" (RM.1) and “climate-related Metrics \& Targets" (MT.1), we report increases of 60\% and 54\% respectively. With regard to the label “climate-related Strategy" (ST.1), we observe an increase of 57 \% in label probability between 2017 and 2021. In the case of “climate-related Governance" (GO.1), we report an increase of 46 \% in label probability between 2017 and 2021. Altogether, these results suggest that reporting on climate-related issues increased after the launch of the TCFD recommendations and, in particular, that climate-related Risk Management reporting exhibited the largest increase, which is consistent with \citet{Ding2022}.

\begin{table}[!ht] \centering
\centering
\begin{scriptsize}
\caption{Mean of label probabilities at category level per financial year}
\label{tab:mean_per_year}
\scalebox{0.6}{
\begin{tabular}{lcccccccccccc}
\hline
                   & 2010 & 2011 & 2012 & 2013 & 2014 & 2015 & 2016 & 2017 & 2018 & 2019 & 2020 & 2021 \\
\hline
Governance         & 0.31 & 0.31 & 0.30 & 0.30 & 0.31 & 0.31 & 0.31 & 0.31 & 0.31 & 0.31 & 0.32 & 0.32 \\
GO.1               & 0.11 & 0.11 & 0.12 & 0.11 & 0.12 & 0.12 & 0.12 & 0.13 & 0.14 & 0.15 & 0.17 & 0.19 \\ \hline
Strategy           & 0.40 & 0.40 & 0.40 & 0.40 & 0.39 & 0.40 & 0.39 & 0.40 & 0.39 & 0.40 & 0.40 & 0.40 \\
ST.1               & 0.12 & 0.12 & 0.12 & 0.12 & 0.12 & 0.13 & 0.13 & 0.14 & 0.15 & 0.17 & 0.20 & 0.22 \\ \hline
Risk management    & 0.23 & 0.23 & 0.24 & 0.23 & 0.24 & 0.23 & 0.23 & 0.23 & 0.23 & 0.23 & 0.24 & 0.24 \\
RM.1               & 0.09 & 0.08 & 0.09 & 0.09 & 0.09 & 0.09 & 0.09 & 0.10 & 0.11 & 0.12 & 0.15 & 0.16 \\ \hline
Metrics \& Targets & 0.31 & 0.31 & 0.31 & 0.31 & 0.31 & 0.32 & 0.31 & 0.32 & 0.32 & 0.32 & 0.33 & 0.34 \\
MT.1               & 0.12 & 0.12 & 0.12 & 0.12 & 0.12 & 0.13 & 0.13 & 0.13 & 0.14 & 0.16 & 0.18 & 0.20\\
\hline
\end{tabular}}
\centering
\begin{flushleft} The table presents the mean of label probabilities (on a scale from 0 to 1) for the general and climate-related labels at category level based on the full sample of 3,355 reports.
\end{flushleft}
\end{scriptsize}
\end{table}

\subsection{Climate-related disclosures by fine-grained TCFD labels}

One way to more precisely assess the quality of climate-related disclosures might be to look not only at the quantity of reporting for each broad TCFD category, but rather to examine more closely corporate reporting on the recommended disclosures within each category. Hence, in the context of the present analysis, we do not consider the overall labels, GO.1, ST.1, RM.1 and MT.1. but rather the fine-grained ones that address specific topics linked to the recommended TCFD disclosures. As highlighted earlier, the TCFD created additional guidance for the financial sector, including banks, insurances, asset managers and asset owners \cite{TCFD2017a}. The additional guidance for banks is aimed in particular at disclosures in Strategy, Risk Management, and Metrics and Targets.

We assume that higher label probability is a proxy for disclosure quality because higher label probability indicates that the semantics of the text sequences are likely to match the semantics of the labels at hand. Thus, when text sequences precisely and explicitly address a topic expressed in a label, they are associated with a high label probability in the context of zero-shot text classification. To some extent, a higher label probability is also an indicator of how much is disclosed about a particular topic, since labeling is associated with a higher probability if a text sequence discusses the topic in question in detail. 

\begin{figure*}[t]
		\centering
			\includegraphics[width=0.70\textwidth]{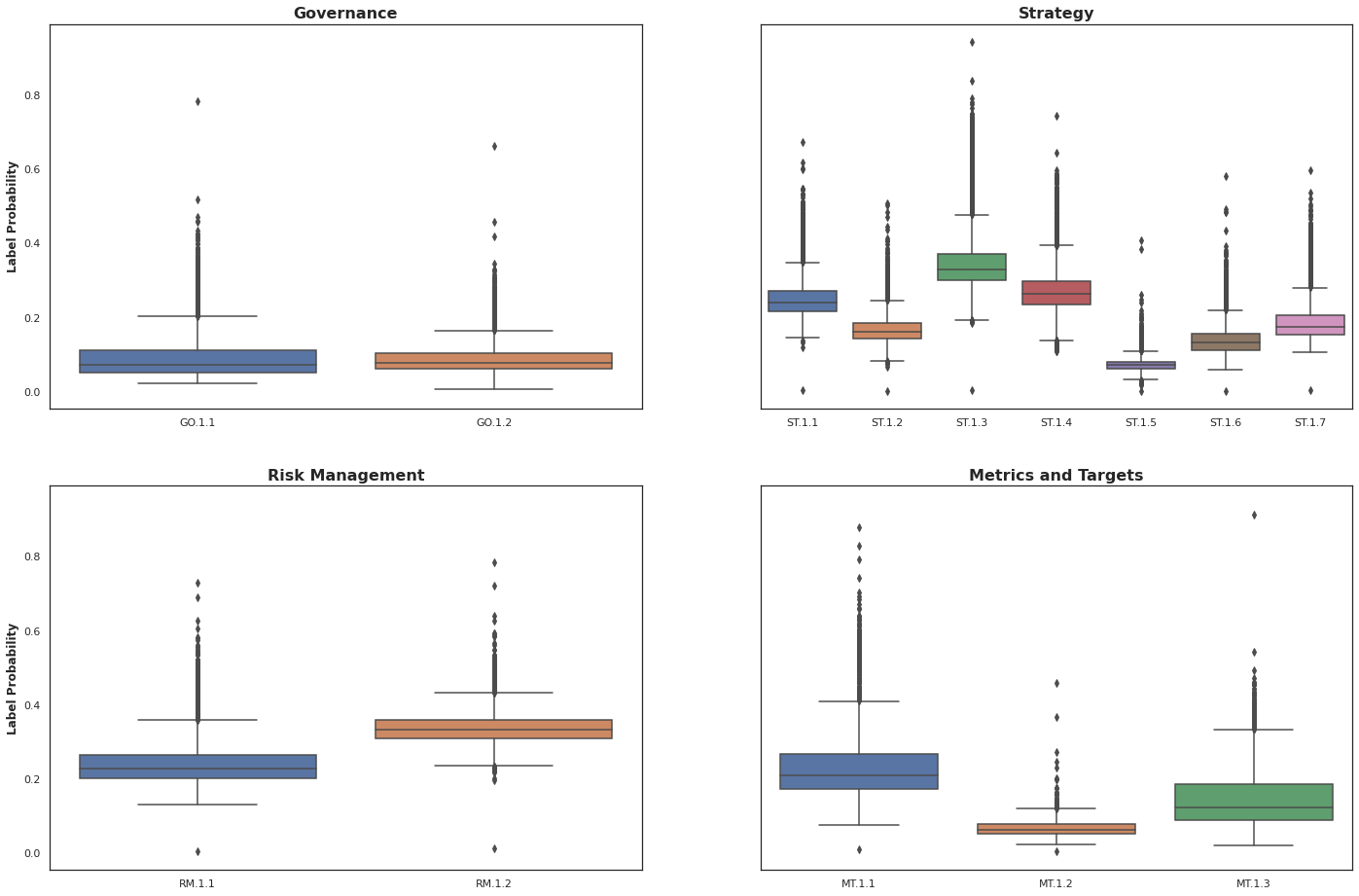}
			\caption{Climate-related disclosures by fine-grained labels}\label{figure:Boxplots2}	
	\end{figure*}

The results of the zero-shot text classification expressed in aggregated average label probabilities are illustrated in Figure \ref{figure:Boxplots2}. As a matter of comparison, Table \ref{tab:ProbsFineLabels} reports the yearly mean label probabilities from 2010 until 2021. The results indicate that there is considerable variation in the extent of disclosure within each TCFD category. In the Governance area, the TCFD-supporting banks appear to report to similar levels on the board's responsibility for overseeing climate-related issues (GO.1.1) and the management's role in assessing and managing climate-related issues (GO.1.2). Both labels achieve a low overall average probability of 9\% and the median is approximately 7\% for GO.1.1 and 8\% for GO.1.2. However, the maximum values for GO.1.1 are higher, implying higher reporting quality on the role of the board in overseeing climate-related issues. This could possibly be due to the fact that our sample includes a broad spectrum of banks, ranging from large to small. Although it is likely that all banks have assigned climate-related responsibilities to management-level positions or committees, it is less likely that smaller banks already conduct extensive oversight of climate-related risks at the board level and are therefore likely to report less information on this item.  

The Strategy area is the most comprehensive and contains several specific recommended disclosures for banks, which explains why there are more labels compared to the other pillars. Turning to the strategy-related labels, we find that reporting on the material financial impact of climate-related issues (ST.1.3) reaches a comparatively high overall average label probability of 35\%. However, as highlighted before, the zero-shot performed less well for this label (F1-score: 0.34), leading to a potentially inflated result. In comparison, the reporting quality appear to be lower for disclosures related to financing and investment for carbon-intensive industries such as fossil fuel industry (ST.1.5) since this label has an overall average probability of only 7\%. This means that of all the text sequences extracted from our full sample of reports and classified by the zero-shot model, the probability that some of them match the semantics of the ST.1.5 label is only 7\%. This is also illustrated in Table \ref{tab:ProbsFineLabels}, where we see that the yearly mean label probabilities of ST.1.5 remain at comparatively low level until 2021 and even after the launch of the TCFD recommendations. Similarly, TCFD-supporting banks also appear to disclose information that is comparatively less likely address climate-related physical risks (ST.1.2) and the use of climate-related scenario models (ST.1.6), as both labels reach overall average probabilities of only 17\% and 14\%, respectively. 

Several aspects could explain these results: For example, the reports in our sample might address only to a limited extent issues linked to the use of scenario analyses or climate-related physical risks since several of the TCFD-supporting banks might not yet have the tools and expertise to perform such analyses or identify such risks. This is consistent with the study on climate-scenarios tools in \citet{Bingler2022}, which finds that only 10 out of 16 existing climate scenario tools allow an assessment of climate-related physical risks. Regarding the low label probability for disclosures related to carbon-intensive industries (ST.1.5), we observe that banks tend to report little information on this topic. In some cases, the text sequences that were attributed this label with a high probability deal with official commitments to stop financing and investments in fossil fuel assets rather than a specific quantification of such investments. In contrast, banks appear to disclose more information on their credit exposure to carbon-related sectors, which is not surprising since this item is more broadly defined and also includes transportation and utilities sectors.

Recent research shows that financing for fossil fuel firms by international banks has not decreased since the Paris Agreement and that international banks continue to provide financing to fossil fuel firms regardless of their stranded asset risk \cite{Beyene2021}. In particular, the analysis performed by \citet{Beyene2021} shows that several TCFD-banks were among the most frequent fossil fuel lenders in the period 2007-2018. In particular, the analysis by the authors reports that JP Morgan (TCFD-member since December 2017), BNP Paribas SA (TCFD-member since June 2017) and Wells Fargo \& Co (TCFD-member since November 2019) to be among the most frequent fossil fuel lenders between 2007 and 2018. Thus, it is also possible that TCFD-banks perform some form of selective disclosure, as postulated by \citet{Bingler2022} and thus omit information out of reputational concerns. At the same time, some banks may not report certain items at all, such as their fossil fuel exposures, because the TCFD recommendations are very broadly worded and do not provide a specific method for measuring this exposure \cite{Beyene2022b}.

In the Risk Management area, we find that, on average, banks reports address less processes for identifying, assessing and managing climate-related risks and integrating them into overall risk management (RM.1.1) than the relationship between climate-related risks and financial risks such as credit risk, market risk, liquidity risk and operational risk (RM.1.2). The median value for RM.1.1 is also lower (23\%) than in the case of RM.1.2 (33\%). However, the probabilities provided by the zero-shot text classification could be slightly inflated compared to the actual reporting since the zero-shot performed less well for RM.1.2 (F1-score: 0.39). As shown previously, several sentence groups achieved a high probability of fitting into RM.1.1 and RM.1.2, which may indicate that the labels and the corresponding recommended disclosures are defined too broadly. In particular, we note that several recommended disclosures within the other pillars (e.g. the role of management in assessing and managing climate-related issues, GO.1.2) are related to risk management issues. 

Under Metrics and Targets, we find that the incorporation of climate-related performance metrics into remuneration policies (MT.1.2) is associated with a lower label probability than metrics related to carbon footprints (MT.1.1) and emissions reduction targets (MT.1.3). The overall average label probability reaches only 7\% compared to 24\% for MT.1.1 and 15\% for MT.1.3. This result is in line with our expectations, as financial institutions are less likely to align their compensation policies with climate-related performance metrics than to more symbolically commit to carbon neutrality goals, even if they might subsequently not meet these commitments, see e.g. \citet{Gibson2019}.

\begin{table}[!ht] \centering
\centering
\caption{Mean of label probabilities for fine-grained labels by financial year}
 \label{tab:ProbsFineLabels}
\begin{scriptsize}
\scalebox{0.7}{
\begin{tabular}{lllllllllllll}
\hline
                     & 2010 & 2011 & 2012 & 2013 & 2014 & 2015 & 2016 & 2017 & 2018 & 2019 & 2020 & 2021 \\
\hline 
GO.1.1               & 0,07 & 0,07 & 0,08 & 0,07 & 0,07 & 0,08 & 0,08 & 0,08 & 0,09 & 0,11 & 0,13 & 0,14 \\
GO.1.2               & 0,08 & 0,07 & 0,08 & 0,07 & 0,08 & 0,08 & 0,08 & 0,08 & 0,09 & 0,10 & 0,11 & 0,12 \\
\hline 
ST.1.1               & 0,24 & 0,24 & 0,24 & 0,23 & 0,24 & 0,24 & 0,24 & 0,24 & 0,25 & 0,26 & 0,28 & 0,28 \\
ST.1.2               & 0,16 & 0,16 & 0,16 & 0,16 & 0,16 & 0,16 & 0,16 & 0,16 & 0,17 & 0,18 & 0,19 & 0,19 \\
ST.1.3               & 0,34 & 0,34 & 0,34 & 0,33 & 0,33 & 0,34 & 0,33 & 0,33 & 0,34 & 0,36 & 0,40 & 0,40 \\
ST.1.4               & 0,25 & 0,25 & 0,26 & 0,25 & 0,26 & 0,26 & 0,27 & 0,26 & 0,27 & 0,28 & 0,31 & 0,32 \\
ST.1.5               & 0,07 & 0,07 & 0,07 & 0,07 & 0,07 & 0,07 & 0,07 & 0,07 & 0,07 & 0,07 & 0,08 & 0,08 \\
ST.1.6               & 0,13 & 0,13 & 0,13 & 0,12 & 0,13 & 0,13 & 0,13 & 0,13 & 0,14 & 0,15 & 0,16 & 0,16 \\
ST.1.7               & 0,18 & 0,17 & 0,18 & 0,17 & 0,17 & 0,18 & 0,18 & 0,18 & 0,19 & 0,20 & 0,22 & 0,22 \\
\hline 
RM.1.1               & 0,22 & 0,22 & 0,23 & 0,22 & 0,23 & 0,23 & 0,23 & 0,23 & 0,24 & 0,25 & 0,27 & 0,28 \\
RM.1.2               & 0,34 & 0,34 & 0,34 & 0,33 & 0,33 & 0,33 & 0,33 & 0,32 & 0,33 & 0,34 & 0,35 & 0,35 \\
\hline 
MT.1.1               & 0,21 & 0,21 & 0,21 & 0,21 & 0,21 & 0,22 & 0,22 & 0,22 & 0,23 & 0,25 & 0,28 & 0,29 \\
MT.1.2               & 0,06 & 0,06 & 0,06 & 0,06 & 0,06 & 0,06 & 0,06 & 0,07 & 0,07 & 0,07 & 0,07 & 0,07 \\
MT.1.3               & 0,12 & 0,12 & 0,13 & 0,12 & 0,13 & 0,13 & 0,14 & 0,14 & 0,15 & 0,16 & 0,18 & 0,19 \\
\hline 
\end{tabular}}
\begin{flushleft} The table presents the descriptive statistics of label probabilities (on a scale from 0 to 1) for the fine-grained TCFD labels based on the full sample of 3,355 reports.
\end{flushleft}
\end{scriptsize}
\end{table}

\section{Conclusion} \label{sec:conclusion}

This paper examines the climate-related disclosures of TCFD-supporting banks using zero-shot text classification as a novel computerized approach for textual analysis. We contribute to the growing literature on voluntary climate-related corporate reporting (e.g., \citet{Friederich2021, Bingler2022, Ding2022}). When combining the TCFD recommendations and the additional financial sector guidance into fine-grained labels, we find large variation in climate-related reporting, not only regarding the broad TCFD categories but also with regard to the underlying recommended disclosure topics. In particular, we find that banks have a lower probability to report on topics such climate-related physical risks, financing and investment for carbon-intensive industries such as fossil fuel industry, the use of climate-related scenario models to analyse the impact of climate-related risks and the incorporation of climate-related performance metrics into remuneration policies. These results indicate that the TCFD-supporting banks in our sample have not yet implemented all the recommendations to the same extent.

Our study entails some limitations, which warrant careful consideration but also may highlight the potential for further research. First, even though we generally find an increase in climate-related reporting following the release of the TCFD recommendations, we cannot say in the context of this study whether banks are actually addressing theses issues in question more intensively. In other words, disclosing more information does not necessarily mean that more internal steps were taken to address the issues at hand.  Second, it could also be that the banks are not reporting about a specific issues because they have not yet implemented internal steps and therefore have nothing to disclose. Third, it might also be that some banks are intentionally omitting information and performing selective disclosure (i.e. greenwashing) out of reputational concerns. Therefore, firms might also seek to improve their public image by controlling the narrative and selectively disclosing information within the TCFD framework that tend to legitimize their actions \cite{Ding2022}. 

Despite these limitations, our study also has important practical implications. In particular, our study highlights the need for precise and specific recommendations when it comes to climate-related disclosure frameworks. The lack of consistent methodologies and specific definitions for recommended disclosure topics appears to lead to wide variation in the scope of reporting. In the case of the banking sector, while the TCFD recommendations are a step in the right direction, the lack of concrete guidance makes it difficult to accurately derive banks' exposure to the fossil fuel sector and ultimately to determine banks' exposure to stranded assets based on the carbon-related metrics disclosed.

\section*{Acknowledgements}
Christoph Funk acknowledges the funding by the German Academic Exchange Service (DAAD) from funds of Federal Ministry for Economic Cooperation (BMZ), SDGnexus Network (Grant No. 57526248), program “exceed - Hochschul\-exzellenz in der Entwicklungszusammenarbeit.” Alix Auzepy acknowledges the funding provided to the project SATISFY, which is part of the research fund Klimaschutz und Finanzwirtschaft (KlimFi) of the German Federal Ministry of Education and Research (BMBF).

\bibliography{references_TCFD} 


\begin{figure*}[t]
		\centering
			\includegraphics[width=0.90\textwidth]{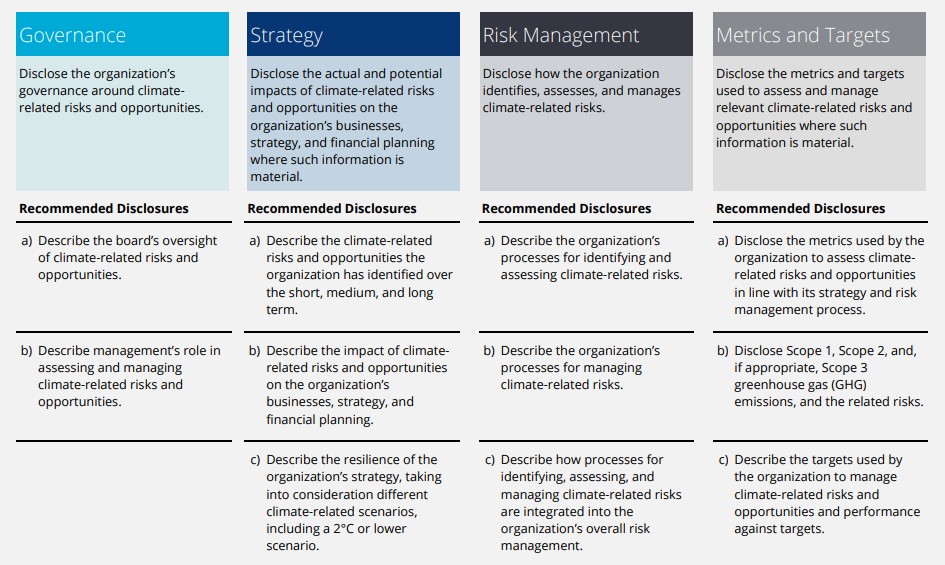}
			\caption{The TCFD disclosure categories and underlying recommended disclosures. \cite{TCFD2017a}}\label{figure:TCFDrecos}	
\end{figure*}

 
\setlength{\tabcolsep}{6pt}
\small
\begin{table*}[t]
    \centering
    \begin{tabular}{ lll}

\hline
\textbf{TCFD Category} & \textbf{Label name} &\textbf{Label Description} \\ \hline
Governance & GO.1. & Climate-related Governance \\
 & GO.1.1 & Board's responsibility for overseeing climate-related issues \\
 & GO.1.2 & Executive management's strategic role related to the assessment and management \\ & & of climate-related issues \\
\hline

Strategy & ST.1. & Climate-related Strategy \\
 & ST.1.1 & Climate-related transition risks such as policy, legal, technology, market and  \\ & & reputation risks   emerging from climate change  \\
 & ST.1.2 & Climate-related physical risks such as acute weather events and chronic shifts in \\ & & weather patterns  \\ 
 & ST.1.3 & Material financial impact of climate-related issues  \\
 & ST.1.4 & Credit exposure to carbon-related sectors   \\
 & ST.1.5 & Financing and investment for carbon-intensive industries such as fossil fuel industry \\
 & ST.1.6 & Use of climate-related scenario models to analyse the impact of climate-related risks  \\
 & ST.1.7 & Resilience of the bank's strategy under different climate-related scenarios \\ 
 \hline

Risk Management & RM.1. & Climate-related Risk Management \\
 & RM.1.1 & Processes to identify, assess and manage climate-related risks and integrate them \\ & & into overall risk management   \\
 & RM.1.2 & Relationship between climate-related risks and financial risks such as credit risk,  \\ & & market risk, liquidity risk and operational risk   \\ 
 \hline

Metrics \& Targets & MT.1. & Climate-related  metrics and targets \\
 & MT.1.1 & Carbon footprint, direct and indirect greenhouse gas emissions   \\
 & MT.1.2 & Incorporation of climate-related performance metrics into remuneration policies   \\ 
 & MT.1.3 & Emissions reduction and carbon neutrality targets    \\ 
 \hline
    \end{tabular}
    \caption{Overview of fine-grained TCFD Labels}
    \label{table2}
\end{table*}    
\begin{figure*}[t]
    \centering
    \includegraphics[width=\textwidth]{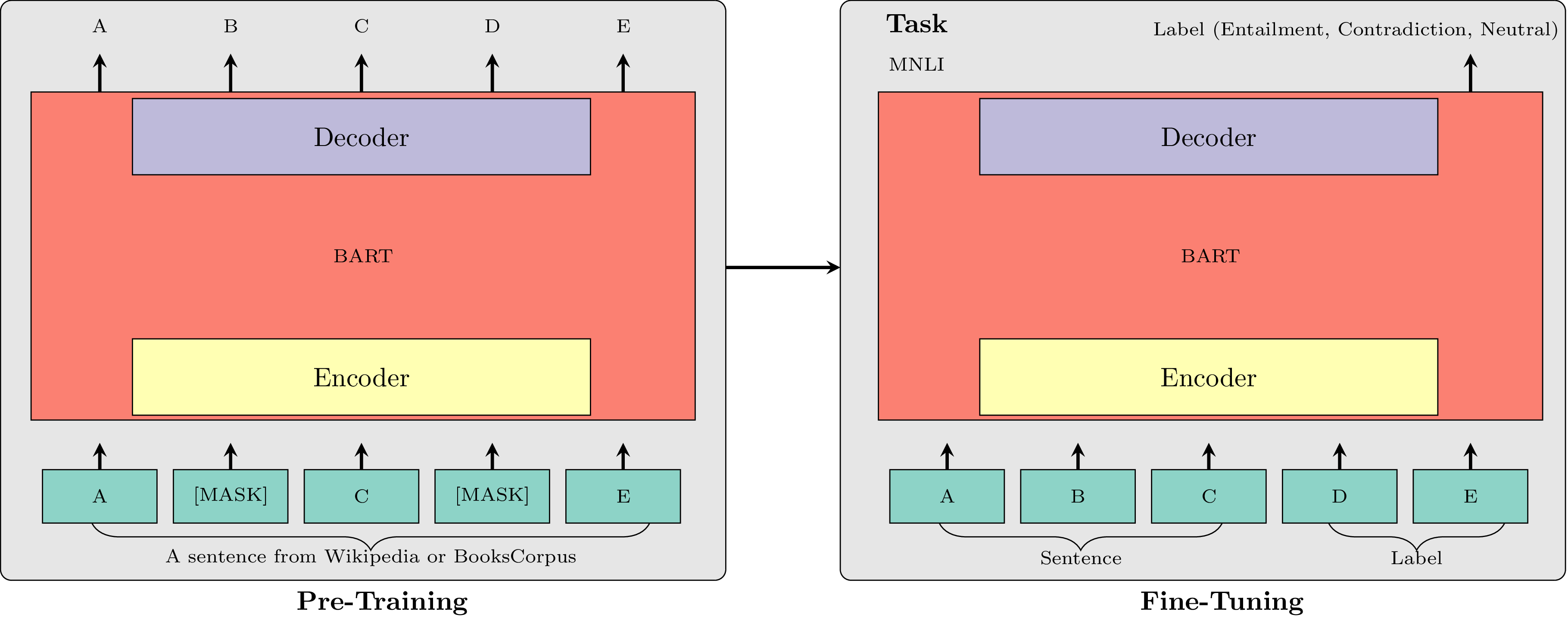}
    \caption{On the left hand side, the BART model is pre-trained on all English Wikipedia articles and the BooksCorpus data set. By masking parts of sentences (\texttt{[MASK]}), the model is trained to learn the semantics and to predict the missing parts. The process is repeated for all sentences in the pre-training data set. On the right hand side, the model is fine-tuned on the MNLI task and returns probabilities for entailment, contradiction and neutral. Own representation based on \citet{Davison.2020}.}
    \label{fig:BART}
\end{figure*}

\begin{figure*}[t]
    \centering
    \includegraphics[width=0.8\textwidth]{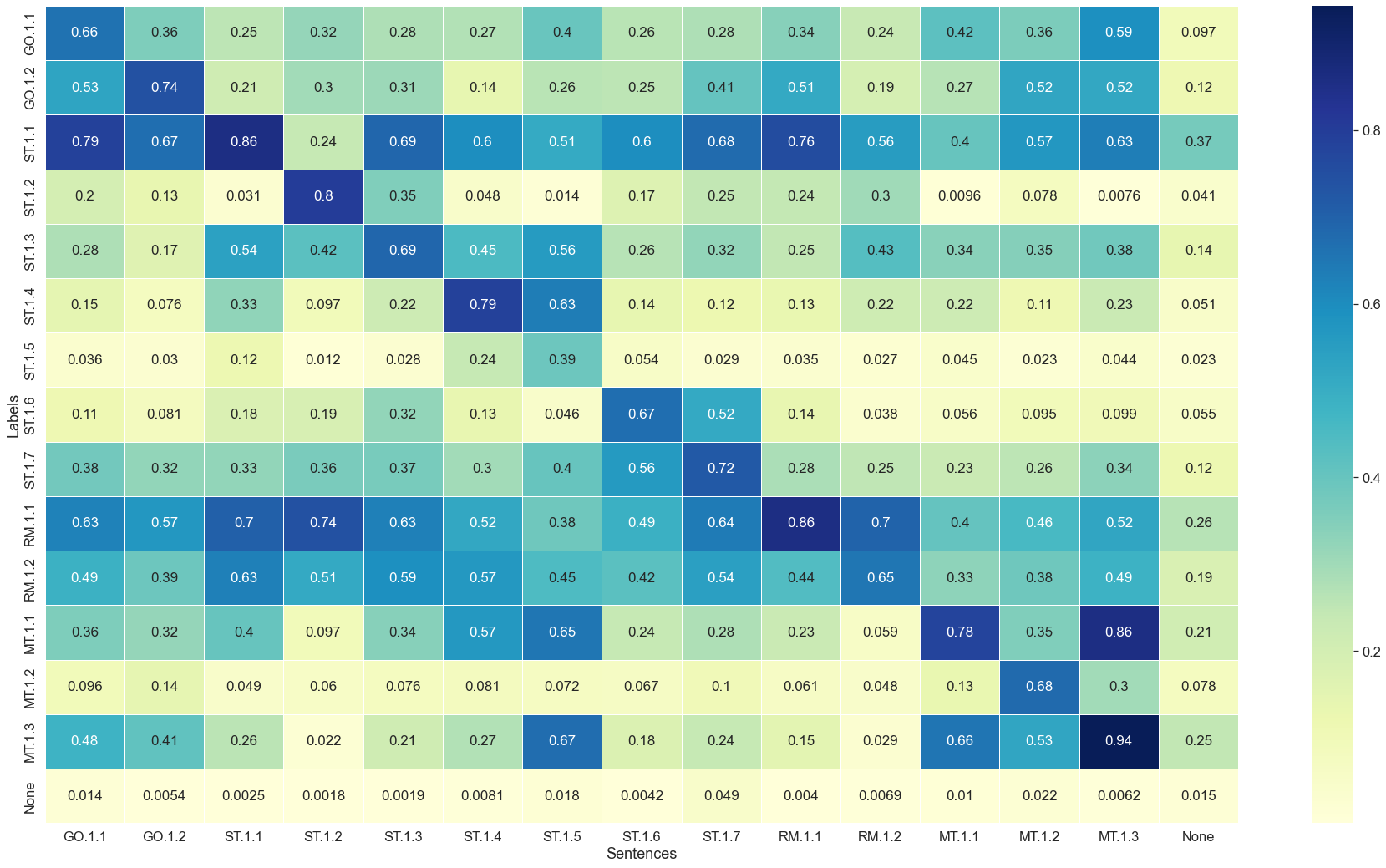}
    \caption{This matrix presents the results of the zero-shot text classification applied to our test data set. The fine-grained TCFD labels are based on the recommended disclosures and the supplemental guidance for the financial sector.}
    \label{fig:our_labels}
\end{figure*}

\end{document}